\documentclass[aps,prl,twocolumn,showpacs]{revtex4}

\usepackage{epsfig}
\usepackage{graphicx}

\begin{document}

\DeclareGraphicsExtensions{.eps,.EPS}

\title{Radio-frequency induced ground state degeneracy in a Chromium Bose-Einstein condensate}
\author{Q. Beaufils, T. Zanon, R. Chicireanu, B. Laburthe-Tolra, E. Mar\'echal, L. Vernac, J.-C. Keller, and O. Gorceix}
\affiliation{Laboratoire de Physique des Lasers, CNRS UMR 7538, Universit\'e Paris 13,
99 Avenue J.-B. Cl\'ement, 93430 Villetaneuse, France}

\begin{abstract}

We study the effect of strong radio-frequency (rf) fields on a
chromium Bose-Einstein condensate (BEC), in a regime where the rf
frequency is much larger than the Larmor frequency. We use the
modification of the Land\'{e} factor by the rf field to bring all
Zeeman states to degeneracy, despite the presence of a static
magnetic field of up to 100 mG. This is demonstrated by analyzing
the trajectories of the atoms under the influence of dressed
magnetic potentials in the strong field regime. We investigate the
problem of adiabaticity of the rf dressing process, and relate it to
how close the dressed states are to degeneracy. Finally, we measure
the lifetime of the rf dressed BECs, and identify a new rf-assisted
two-body loss process induced by dipole-dipole interactions.
\end{abstract}


\date{\today}

\maketitle

When all magnetic substates $\left|m\right\rangle$ of an atomic
species ground state are nearly degenerate, it becomes possible to
study new features related to the vectorial nature of the spin in
the ground state of multicomponent either Bose condensed \cite{Ho},
or Fermi degenerate optically trapped gases \cite{azaria}. These
systems are known as spinor quantum gases. To explore these new
features, it is important that differences in interaction energies
between different total spin states are larger than their relative
Zeeman energy, which requires magnetically shielded environments.

Up to now experiments on spin-1 \cite{spinorF1} and spin-2
\cite{spinorF2} spinor condensates were typically performed starting
with atoms in the $\left|m=0\right\rangle$ magnetic state, with
emphasis on spin dynamics and coherent oscillations between the spin
components. Since spin dynamics is driven by spin exchange
collisions, which do not modify the total spin angular momentum, one
therefore works in a subspace insensitive to first order Zeeman
effect, but the spinor ground state is not obtained
\cite{stamper-kurn}. A subspace insensitive to magnetic fields to
first order is also used for quantum computing purposes with cold
atoms in optical lattices, to reduce decoherence during quantum gate
operations \cite{treybloch}. More generally, a very accurate control
of the magnetic fields is required for precision measurements (e.g.
atomic clocks use both magnetic shielding and a transition
insensitive to the Zeeman effect to first order).

To ease the constraints on magnetic field control, we suggest to use
strong off resonant linearly polarized rf fields, to bring all
Zeeman states to degeneracy despite a non-zero magnetic field. We
demonstrate this idea by sending strong rf fields to optically
trapped Bose condensed chromium atoms. We analyze the trajectories
of atoms in dressed magnetic potentials and we show that, as
expected from \cite{Haroche}, the Land\'{e} factor is modifed, and
can even be set to zero. At this point, all Zeeman states are
degenerate. We show that the adiabaticity criterion for ramping up
the rf power strongly depends on such degeneracy. Finally, we
discuss inelastic losses measured in the dressed sample, and
attribute them to an exoenergetic rf-assisted dipolar coupling to
higher partial waves.

Before describing our experimental results, let us give a physical
insight into the modification of the atoms eigen-energies by the rf
field. As shown in \cite{Haroche} using first order perturbation
theory, when the rf frequency $\omega$ is much larger than the
Larmor frequency $\omega_{\perp}$ the Land\'{e} factor $g_J$
perpendicular to the rf field axis is modified by the rf dressing of
the atom and is given by:

\begin{equation}
g_J (\Omega) = g_J J_0\left(\frac{\Omega}{\omega}\right)
\label{landeeq}
\end{equation}

where $\Omega= g_J \mu_B B_{rf}/\hbar$ is the Rabi angular
frequency, $\mu_B$ the Bohr magneton, and $J_0$ is the zero order
Bessel function. As a result, the eigenenergies of the different
$\left|m\right\rangle$ states dressed by rf, in presence of a static
magnetic field read:

\begin{equation}
E_m = m \mu_B g_J \sqrt{\left(B_{\perp}
J_0\left(\frac{\Omega}{\omega}\right)\right)^2+B_{//}^2}
\label{landeeq2}
\end{equation}
where $B_{//}$ and $B_{\perp}$ stand for the components parallel and
perpendicular to the rf field.

When $\Omega$ is such that the Bessel function is zero, atoms are
insensitive to transverse magnetic fields. We have derived a
convenient picture of this effect from the classical dynamical
equation of a spin in presence of a rf field $\vec{B}(t)=B_{rf}
\cos(\omega t) \vec{z}$, which reads $d \vec{\mu}/dt = \frac{g_J
\mu_B}{\hbar}\vec{\mu}\times\vec{B}(t)$. This equation has an
analytical solution $\propto \cos(\frac{\Omega}{\omega} \sin(\omega
t) ) $ which, when time-averaged, leads to $\left\langle
\mu_x\right\rangle$ , $\left\langle \mu_y\right\rangle \propto J_0
\left( \frac{\Omega}{\omega} \right)$, where $\left\langle
\mu_x\right\rangle$ and  $\left\langle \mu_y\right\rangle$ are the
time averaged values of $\vec{\mu}$ perpendicular to the rf field.
In presence of a small bias field $B_{\perp}\vec{x}$ perpendicular
to $\vec{B}_{rf}$ the average energy of a classical spin is
$\left\langle\mu_x\right\rangle B_{\perp} \propto g_J (\Omega) \mu_B
B_{\perp}$. In this picture, the effect of rf on $g_J$ can be seen
as resulting from a sinusoidal modulation of the frequency of
precession of the atoms. As in many other systems (frequency
modulation of a laser, sinusoidal diffraction of light or
matter-waves, modulation of the depth of an optical lattice
\cite{Arimondo}, modulation of the eigenenergies of Rydberg atoms
\cite{Pillet}), this results in the occurrence of Bessel functions.

The rf-renormalization of $g_J$ described in \cite{Haroche} was
first probed using microwave spectroscopy \cite{schermann} and
through the modification of spin-exchange collisions between Rb and
Cs atoms \cite{Haroche2}. In both these experiments, the magnetic
fields were in the micro-Gauss range. Here, we observe the reduction
of magnetic forces on an optically trapped chromium Bose-Einstein
condensate, in presence of magnetic fields up to 100 mG. We therefore greatly reduce the
sensitivity of atoms to magnetic fields up to a value easily
controlled by experimentalists, even without magnetic shielding.

\begin{figure}
\centering
\includegraphics[width= 2.8 in]{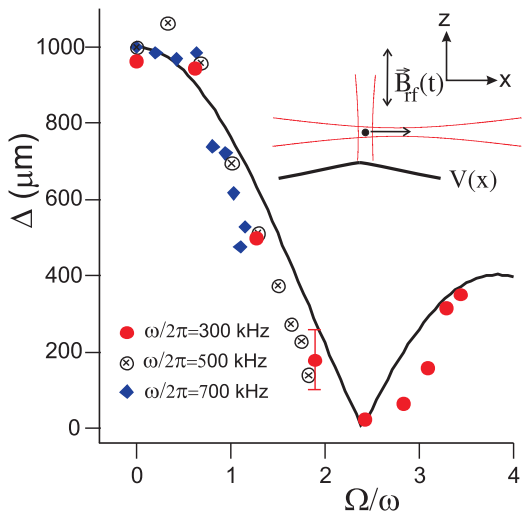}
\caption{\setlength{\baselineskip}{6pt} {\protect\scriptsize
Displacement $\Delta$ of the dressed BEC after 35 ms of drift in the
horizontal trap, relative to the displacement of $\left|
m=0\right\rangle$ atoms under similar conditions, as a function of
rf power, for three different rf frequencies. Line: $\Delta_{max}
\times |J_0\left(\frac{\Omega}{\omega}\right)|$. The relative
uncertainty on the horizontal axis is about 10 \%. The error bar
gives (as in Fig 2) the typical statistical uncertainty. Insert:
sketch of the experimental set-up. The BEC is formed at the
intersection of two dipole traps. When the vertical one is switched
off, atoms drift in the horizontal trap and are accelerated by the
magnetic potential V(x).}} \label{deplacementBEC}
\end{figure}

Our recipe to produce $^{52}$Cr Bose-Einstein condensates is
described in \cite{BECpaper}. Forced evaporation is performed in a
crossed optical dipole trap, and BECs are produced with typically 10
000 atoms in the absolute ground state
$\left|S=3,m=-3\right\rangle$, in about 14 s. At this stage, the
magnetic field is 2.3 G.

After BEC has been reached, we adiabatically recompress the optical
dipole trap (then, the chemical potential is about 4 kHz), and
reduce the magnetic field at the BEC position to reach a Larmor
frequency of 85 kHz. We characterized the magnetic field at the
atoms position to a precision of 2 mG by rf spectroscopy. In
addition, we also measure a magnetic field gradient of $b'=$ 0.25
G/cm along the axis of the horizontal dipole trap. After the
magnetic field has reached its final value, atoms are released into
the horizontal optical trap, by suddenly removing the vertical
trapping beam. The atoms are then accelerated by the magnetic field
gradient $b'$. Atoms in the $\left|m=-3\right\rangle$ state
experience an anti-confining potential $V(x)= -m g_J \mu_B b'
\left|x\right|$, and they are expelled from the center of the trap.
Due to the radial confinement of the dipole trap, the motion of the
atoms is channeled in one direction. In fact, the BEC is not
produced exactly at the waist of the trapping laser, so that atoms
also experience a force due to the gradient of the dipole
longitudinal potential (see insert in Fig \ref{deplacementBEC}). We
therefore measure the displacement of the atoms, relative to the
displacement of atoms in $\left|m=0\right\rangle$ under similar
conditions. This additional longitudinal displacement $\Delta (t)$
of the BEC after a time $t$ provides a measurement of $g_J(\Omega)$ since $\Delta(t) = \frac{1}{2}\frac{m g_J\left(\Omega\right) \mu_B b'}{M}
t^2$, where $M$ is the atom mass. Using a BEC enables us to precisely
measure $\Delta (t)$ for "long" delay without being disturbed by
substantial expansion of the cloud.

Radio-frequency fields are applied to the atoms using a 150 W rf
amplifier driving a 8 cm diameter, 8-turn coil, located 4 cm away
from the atoms. The rf frequency is larger than all Larmor
frequencies at any given position of the atomic cloud trajectory.
When a sufficiently strong rf field is applied, the trajectory of
the atoms is modified as they travel through rf-dressed adiabatic
potentials. Fig \ref{deplacementBEC} represents $\Delta$(35 ms) as a
function of $\Omega / \omega$, for three different frequencies
$\omega$. For this experiment, we ramped the rf in 1 ms up to its
final value $\Omega $. The change in position as the rf power is
modified is a signature of the modification of $g_J$. For each value
of $\omega$, the Rabi frequency of the atoms was precisely
calibrated by measuring Rabi oscillations in a magnetic field
$\vec{B_0}\parallel \vec{x}$ with a Larmor frequency $\omega_0 =
\omega$. All data points lie close to the same universal curve
corresponding, for $\Omega/\omega <2.4$, to eq. (\ref{landeeq}).
There is no adjustable parameter on the horizontal axis and the
amplitude of the Bessel function is set by $\Delta_{max}$, the
displacement of the atoms when the rf is off. Up to $\Omega/\omega
=2.4$ , at which point $g_J(\Omega)$=0, our data is therefore
consistent with theoretical predictions.

\begin{figure}
\centering
\includegraphics[width= 3 in]{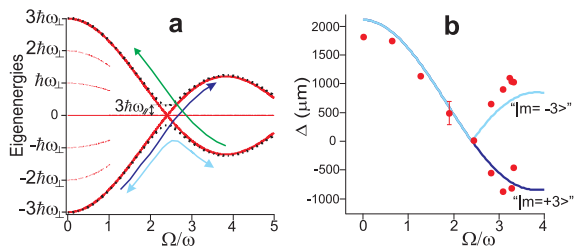}
\caption{\setlength{\baselineskip}{6pt} {\protect\scriptsize (a)
Sketch for adiabaticity issues. We plot the eigenenergies of states
adiabatically connected to the different Zeeman states, when the
longitudinal field $B_{//}$ is null (red), or not (dotted black).
The arrows represent: in light blue the dressing and undressing
corresponding to fully adiabatic process, in dark blue (green) a
fully diabatic dressing (undressing) process. (b) Similar results
than in Fig 1 (with $\omega/2\pi=$300 kHz), except that the rf power
is ramped up in 20 $\mu$s instead of 1 ms. For $\Omega/\omega >2.4$
many dressed states are populated and we only represent the
displacement of the extreme ones.}}

\label{Eigenenergies}
\end{figure}

For $\Omega/\omega >2.4$, the agreement breaks down. Instead of
changing sign, as predicted by eq. (\ref{landeeq}), $g_J$ remains
positive, and raises again. To understand this issue, we refer to
Fig \ref{Eigenenergies}a, where the eigenenergies $E_m$ (see eq.
(\ref{landeeq2})) of the dressed states are qualitatively
represented. As expected from eq. (\ref{landeeq}), when the Bessel
function approaches zero, all eigenstates get nearly degenerate.
There is nevertheless an avoided crossing associated to the presence
of a small bias field component $B_{//} $ parallel to the rf field.
The fact that we are not able to reverse magnetic forces on the
atoms, evidenced in Fig \ref{deplacementBEC}, indicates that we are
adiabatic while reaching this avoided crossing, whereas one needs to
be fully diabatic to follow the Bessel-function curve. We therefore
expect that the experimental points in Fig \ref{deplacementBEC}
should follow the absolute value of the Bessel function, as we
observed.

\begin{figure}
\centering

\includegraphics[width= 2.8 in]{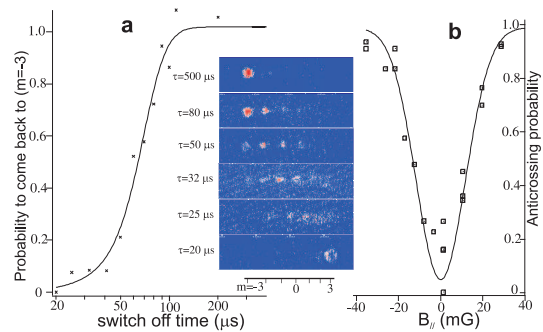}
\caption{\setlength{\baselineskip}{6pt} {\protect\scriptsize (a)
Measurement of the probability of recovering the initial $\left|
m=-3\right\rangle$ state after rf dressing, as a function of the
switch-off time $\tau$ for $\Omega/\omega=2.8$. Absorption pictures
after 35 ms of expansion in the horizontal dipole trap, in presence
of the magnetic field gradient, for different values of $\tau$. (b)
Probability for an adiabatic dressing process as a function of
$B_{//}$. The rf is switched on in  20 $\mu s$ and applied for 45 ms
with $\Omega/\omega=3.25$. Lines are guides for the eyes.}}
\label{adiabatique}
\end{figure}

To improve our understanding of the adiabaticity issues, we
performed additional experiments. We raise the rf power in a much
shorter  time (20 $\mu$s), and let the dressed atoms expand in the
magnetic gradient. The atoms remain in one eigenstate for
$\Omega/\omega<2.4$, but when the avoided crossing is reached, the
BEC is projected on a superposition of all dressed states. We plot
in Fig \ref{Eigenenergies}b the position of the two extreme dressed
states after 45 ms of drift in the horizontal trap. One follows the
Bessel function, the other its absolute value. This indicates that
for a 20 $\mu$s ramp up time, the crossing of the point at 2.4 is
not adiabatic. To be more quantitative, we performed the following
experiments, described in Fig \ref{adiabatique}. We apply the rf
field to the atoms initially polarized in $\left| m=-3\right\rangle$
and study how the switch off time impact on the probability of
recovering the initial state. The rf power is ramped up in 1 ms,
stays up for $(1-\tau)$ ms, and ramped down in $\tau$. We switch off
the vertical trapping beam, and perform a Stern-Gerlach experiment:
the atoms expand in the horizontal dipole trap, and the magnetic
field gradient separates the different $\left|m\right\rangle$
states. We plot in Fig \ref{adiabatique}a the probability of
remaining in the $\left| m=-3\right\rangle$ state after the rf ramp,
as a function of $\tau$, at a rf frequency of 300 kHz. If $\tau > $
100 $\mu$s, the atoms come back to the initial state, showing that
the process is adiabatic, as represented by the light blue arrow in
Fig \ref{Eigenenergies}a. When $\tau$ is small, instead of coming
back to the initial $\left| m=-3\right\rangle$ state, we populate
mostly $\left| m=+3\right\rangle$, as illustrated in the false color
pictures of Fig \ref{adiabatique}. This indicates that on this
timescale, crossing the $\Omega / \omega =2.4$ point is diabatic.
This process is represented by the green arrow in Fig
\ref{Eigenenergies}a.

In Fig \ref{adiabatique}b, we show the influence of $B_{//}$ on the
adiabaticity time scale. The rf is ramped up in 20 $\mu$s to reach
$\Omega/\omega=3.25$ and we measure for different values of $B_{//}$
the population of the atoms following the adiabatic trajectory
(upper branch in Fig \ref{Eigenenergies}b). We can qualitatively
interpret the influence of $B_{//}$ on the adiabaticity time scale
using a Landau-Zener criterion $\frac{dE}{\hbar dt} \approx
\omega_{//}^2$ where $\omega_{//}$ is the Larmor angular frequency
associated with $B_{//} $, $\frac{dE}{dt} \approx \frac{3\hbar
\omega_{\perp}}{\Delta t}$ and $\Delta t=$20 $\mu s$ the rf raising
time. A good adiabaticity is then expected for $B_{//}>$ 20 mG ,
which is consistent with our measurements.

Breakdown of adiabaticity as illustrated in Fig \ref{Eigenenergies}
and \ref{adiabatique} is thus a signature of all dressed eigenstates
getting close to degeneracy. On the other hand, in the prospect of
using this rf dressing technique to reach the ground state of spinor
systems, the fact that increasing the rf power sufficiently slowly
is reversible (see Fig \ref{adiabatique}) is important: to remain in
the ground state of the many-body system, it is important to make
sure that dressing and undressing of the atoms are indeed adiabatic,
at least in the single particle limit. In practice, we do observe
that a polarized BEC is recovered with no substantial heating after
having interacted with the rf.

\begin{figure}
\centering
\includegraphics[width= 2.8 in]{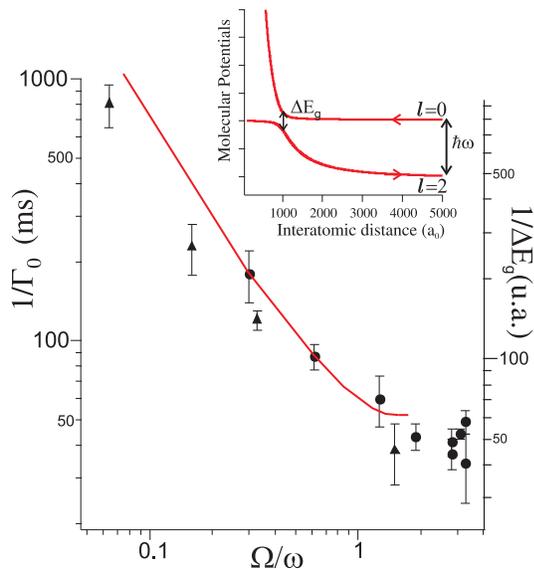}
\caption{\setlength{\baselineskip}{6pt} {\protect\scriptsize Left:
inverse of the experimental BEC decay rate at short time for
different rf powers at rf frequencies of 300 kHz (filled circles) or
500 kHz (triangles). Right (solid line): inverse of the calculated
gap $\Delta$E$_g$. Insert: molecular potentials of two adjacent
manifolds with $l=2$ and $l=0$ to illustrate our model for losses.
}} \label{lifetime}
\end{figure}

In a spinor BEC one has to consider the interaction between the
particles. In this prospect, we first investigate the question of
the collisional stability of the BEC when dressed by rf. We
performed measurements of the BEC lifetime in the crossed optical
dipole trap. We observe density-dependent non-exponential decay, and
we report on Fig \ref{lifetime} the inverse of the decay rate at
short time, $\Gamma_{0}$, as a function of the rf power. Although
lifetimes as small as 50 ms are obtained, the trap frequencies are
on the order of 300 Hz, and the chemical potential, 4 kHz. This
insures thermal equilibrium as the dressed BEC decays.

As the atoms are in the lowest state of energy of their manifold of
rf-dressed states, the inelastic process necessarily implies a
coupling to a lower manifold. Such a coupling can only result from
the spin-dependent part of the inter-particle potential, which is
necessarily the magnetic dipole-dipole potential since there are no
hyperfine interactions for $^{52}$Cr. In order to get an insight on
the loss mechanism, we numerically solved the problem of two spin
1/2 atoms in $\left|m=-1/2\right\rangle$ dressed by rf in the strong
field regime, in presence of dipole-dipole interactions. Such atoms
are colliding in the state $\left| S=1, m_S = -1, l=0, m_l=0
\right\rangle$ - where $l$ defines the incoming partial wave -
belonging to a given manifold. We find that an avoided crossing
opens with rf power between the molecular adiabatic potentials
corresponding to this state, and to the state $\left| S=1, m_S = -1,
l'=2, m_l'=-1 \right\rangle$ belonging to the nearest lower
manifold. The avoided crossing occurs at a distance $R_c$ such that
the centrifugal barrier energy $\frac{l'(l'+1) \hbar^2}{2 \mu R_c^2}
\approx \hbar \omega$, with $\mu=M/2$. Typically, $R_c \approx $900
$a_0$, and the attractive molecular potential does not come into
play (see insert of Fig \ref{lifetime}). The gap $\Delta$E$_g$ is
proportional to the dipole-dipole coupling strength $V_d$, and for
large rf power, $\Delta$E$_g\approx V_d(R_c)$. Pairs of atoms
transferred in the lower manifold by this mechanism acquire a
kinetic energy $\hbar \omega$ and are expelled from the trap.

This mechanism reminds of dipolar relaxation \textit{without rf},
for atoms in the stretched state of \textit{maximum} energy in a
static magnetic field $B_{eq}$. Then an avoided crossing with a gap
$\approx V_d(R_d)$ opens between the potentials of $\left| S=1, m_S
= 1, l=0, m_l=0 \right\rangle$ and $\left| S=1, m_S = 0, l'=2,
m_l'=1 \right\rangle$ at a distance $R_d$ such that $\frac{l'(l'+1)
\hbar^2}{2 \mu R_d^2}= g_J \mu_B B_{eq}$. Therefore, the two-body
loss parameter that we expect for this rf-assisted mechanism is on
the same order of magnitude as $K_2^{rel}$, the two-body loss
parameter for dipolar relaxation in a static field $B_{eq} \approx
\frac{\hbar \omega}{g_J \mu_B }$. Given the known value for
$K_2^{rel}$ \cite{Hensler}, and our typical BEC density, a typical
lifetime of a few tens of ms is expected at large rf power. Although
a full quantitative analysis is beyond the scope of this paper, we
represent in Fig \ref{lifetime} the evolution of the gap we have
calculated as a function of rf power. A saturation at $\Omega/\omega
\approx 2$ is obtained, as for our the measured losses. At larger
power, $\Delta$E$_g$ starts decreasing again, but other gaps open,
coupling the initial state to even lower manifolds.

Controlling the magnetic state degeneracy would be very useful for
many applications. We have already stressed the relevance of this
issue to spinor physics. The use of strong rf fields to achieve this
goal reduces significantly the BEC lifetime in the case of chromium
due to strong dipole-dipole interactions. However, we expect much
larger life times for atomic species with a smaller magnetic moment.
For some applications, a complete 3D effective magnetic shielding
may be required. We are currently theoretically investigating this
issue by considering the interaction of atoms with two perpendicular
rf fields at two different frequencies.

Acknowledgements: LPL is Unit\'e Mixte (UMR 7538) of CNRS and of
Universit\'e Paris 13. We acknowledge financial support from
Minist\`{e}re de l'Enseignement Sup\'{e}rieur et de la Recherche and
IFRAF (Institut Francilien de Recherche sur les Atomes Froids). We
thank Pr. Ennio Arimondo for stimulating discussions.

\end{document}